\documentclass[aps,prl,twocolumn,showpacs,preprintnumbers,floatfix]{revtex4}
\usepackage{graphicx}
\usepackage{amsmath}
\usepackage{amssymb}
\usepackage{epsfig}
\usepackage{color}

\begin{document}

\title{Excitation and control of chirped nonlinear ion acoustic waves}
\author{L. Friedland }
\affiliation{Racah Institute of Physics, Hebrew University of Jerusalem, Jerusalem 91904,
Israel}
\author{A. G. Shagalov}
\affiliation{Institute of Metal Physics, Ekaterinburg 620219, Russian Federation}

\begin{abstract}
Large amplitude ion acoustic waves are excited and controlled by a chirped
frequency driving perturbation. The process involves capturing into
autoresonance (a continuous nonlinear synchronization) with the drive by
passage through the linear resonance in the problem. The transition to
autoresonance has a sharp threshold on the driving amplitude. The theory of
this transition is developed beyond the Korteweg-de-Vries limit by using the
Whitham's averaged variational principle within the water bag model and
compared with Vlasov-Poisson simulations.
\end{abstract}

\maketitle

Waves in continuous media can be excited by a variety of processes involving
resonant wave interactions. The approach requires phase matching and the
examples range from optical fiber or superconducting
parametric amplifiers \cite{Agrawal,Hatridge} to the formation of electrostatic waves in
stimulated Raman (SRS) and Brilluin (SBS) scattering in laser plasma
interactions \cite{Kruer} and more. The nonlinearity, as well as variation
of the parameters of the background medium shift the frequencies and the
wave vectors of the excited waves and, thus, tend to destroy the phase
matching in resonant interactions leading to saturation of the excitation
process. Nevertheless, under certain conditions, the nonlinearity and the
variation of system parameters may work in tandem to dynamically preserve
the phase matching. This phenomenon is called autoresonance \cite%
{Scholarpedia}. It was studied in many applications, such as particle
accelerators \cite{Livingston}, fluids \cite{Fluids}, planetary dynamics
\cite{Malhotra}, atomic systems \cite{Atomic}, optics \cite{Optics} and
more. In plasmas, the autoresonance idea was used for generation of large
amplitude plasma waves in beat-wave accelerators \cite{BeatWave}, excitation
of the diocotron modes in pure electron plasmas \cite{Diocotron}, in recent
experiments at CERN on formation of cold antihydrogen atoms \cite{CERN}, as
well as in the SRS and SBS theory \cite{SRS,SBS}.
\begin{figure}[bp]
\includegraphics[width=8.5cm]{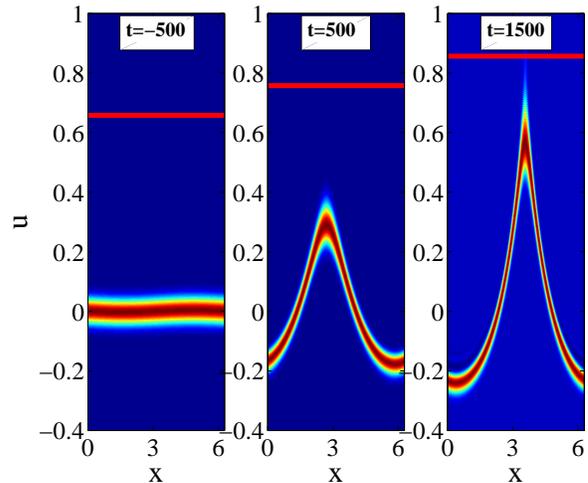}
\caption{(color online) The autoresonant ion distribution in the driving
wave frame at three successive times. The horizontal red lines show the
location of the driving phase velocity.}
\label{Distribution}
\end{figure}

The ion acoustic waves are low frequency longitudinal plasma density
oscillations. They were predicted by Tonks and Langmuir \cite{Tonks} in 1929
on the bases of the fluid theory and observed experimentally by Revans in
1933 \cite{Revans}. Since then, this important branch of plasma waves was
studied in a variety of different contexts, such as laser-plasma
interactions \cite{Kruer} and dusty \cite{Dusty}, ionospheric \cite%
{Ionospheric}, and ultra cold \cite{UltraCold} plasmas. The SBS is one of
the most important resonant three-wave interaction processes in laser fusion
plasmas involving ion acoustic waves \cite{Kruer,Lindl}. It describes the
decay of the incident high power laser radiation (the pump) in the plasma
into the scattered electromagnetic wave and an ion acoustic wave. The
process is one of the causes of depleting and redirecting the incident laser
flux. Despite its importance, the theoretical understanding of this
phenomenon is still incomplete for plasmas characteristic of many present
experiments, the reason being the complexity involving such factors as the
nonlinearity \cite{Casanova}, plasma nonuniformity and time dependence \cite%
{Drake}, and the effects of resonant particles \cite{Cohen}. All these
factors affect the phase matching condition between the waves, while the
kinetic effects lead to Landau damping, resonant trapping of plasma
particles and consequent saturation. Can one preserve the phase matching and
avoid saturation of the ion acoustic wave by autoresonance in the system?
This work is devoted to studying this question in the small ion temperature
limit $T_{i}/T_{e}\ll 1$ limit. Previous related work used the long wave
length fluid-type, driven KdV model \cite{Friedland98,Maximov}. We will
advance the analysis by allowing for arbitrary nonlinearity, longer wave
length ($k\lambda _{D}\sim O(1)$) and finite ion temperature.
\begin{figure}[tp]
\includegraphics[width=9cm]{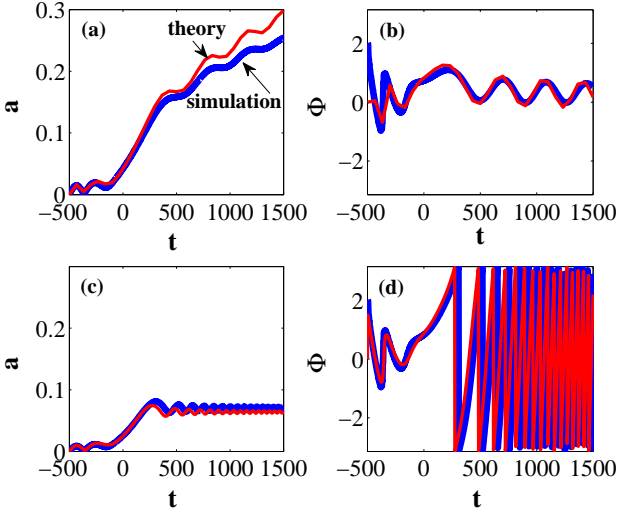}
\caption{(color online) The evolution of the amplitude $a$ and phase
mismatch $\Phi $ above ((a), (b)) and below ((c), (d)) the threshold. The
thin (red) lines show Vlasov-Poisson simulations, the thick (blue) lines the
theoretical results.}
\label{Amplitude&Phase}
\end{figure}
This work is motivated by numerical simulations of the following
one-dimensional Vlasov-Poisson system describing a driven ion acoustic wave%
\begin{equation}
f_{t}+uf_{x}-\varphi _{x}f_{u}=0,\text{ }\varphi _{xx}=\exp (\varphi
+\varphi _{d})-\int fdu\text{ }  \label{VL}
\end{equation}%
Here $f$ and $\varphi $ are the ion distribution and the electric potential
and $\varphi _{d}=\varepsilon \cos \theta _{d}$, where $\theta _{d}=kx-\int
\omega _{d}dt$ is a small amplitude wave-like driving potential, having a
slowly varying frequency $\omega _{d}(t)$. All dependent and independent
variables in (\ref{VL}) are dimensionless, such that the position, time, and
velocity are rescaled with respect to the Debye length $\lambda
_{D}=u_{e}/\omega _{p}$, the modified inverse plasma frequency $%
(m_{i}/m_{e})^{1/2}\omega _{p}^{-1}$, and the modified electron thermal
velocity $(m_{e}/m_{i})^{1/2}u_{e}$. The distribution function and the
potentials are rescaled with respect to $(m_{i}/m_{e})^{1/2}n_{0}/u_{e}$,
and $k_{B}T_{e}/e$, respectively (here $k_{B}T_{e,i}=m_{e,i}u_{e,i}^{2}$).
We assume that the electrons are Boltzmanian in the combined driven and
driving potentials. We also assume spatial periodicity of period $2\pi /k$
associated with the driving wave and solve the time evolution problem,
subject to the simplest initial equilibrium: $\varphi (x,0)=0$ and $%
f(u,x,0)=(2\pi \sigma ^{2})^{-1/2}\exp (-u^{2}/2\sigma ^{2})$, where $\sigma
^{2}=T_{i}/T_{e}$. Note that $\sigma $ and the driving parameters fully
define our rescaled, dimensionless problem. We applied our Vlasov code \cite%
{Lazar111} for solving this problem numerically and show the results of the
simulations in the driving wave frame in Fig. 1 for $\sigma =0.03$, i.e. $%
T_{i}\ll T_{e}$. We increase the driving frequency, $\omega _{d}=\omega
_{0}+\alpha t$ , and use parameters $k=1$, $\omega _{0}=0.66$, $\alpha
=0.0001$, and $\varepsilon =0.0022$. Our simulations show that after the
driving frequency passes the linear ion acoustic frequency, $\omega
_{a}=k(1+k^{2})^{-1/2}$, the system phase-locks to the drive and one
observes the formation of a growing amplitude autoresonant deformation of
the ion distribution in the figure. The associated density perturbation
comprises a continuously phase-locked, growing amplitude ion acoustic wave.
Note that at all stages in this example, the driving phase velocity (its
location is indicated by the straight red line in the figure) is well
outside the ion distribution and, thus, the effect of resonant particles is
negligible, until the final stage at $t=1500$, when some resonant particles
can be seen in the simulations. We found that the phase-locking was lost
beyond this stage. Figures 2a,b show the time evolution of the amplitude $\ a
$ of the first harmonic of the electric potential $\varphi $ of the ion
acoustic wave and the phase mismatch $\Phi $ between the driven and driving
waves as obtained in the simulations (full lines) and theory (dotted lines)
presented below. Importantly, we also found that the autoresonant excitation
as seen in Fig. 1 took place only if the driving amplitude exceeded a
threshold $\varepsilon _{th}$ ($\varepsilon _{th}=0.0017$ in our example).
Below the threshold, the excitation saturates (see Figs. 2c,d for $%
\varepsilon =0.0013$) and the phase-locking discontinues.
\begin{figure}[bp]
\includegraphics[width=9cm]{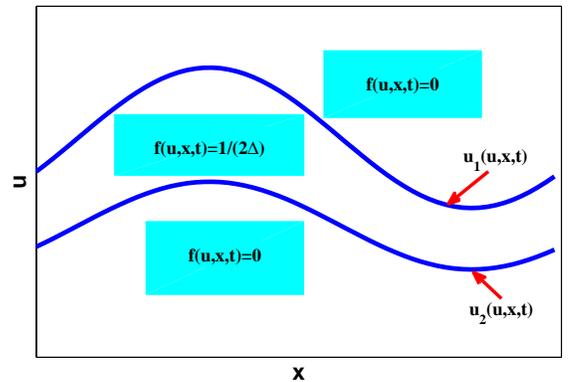}
\caption{(color online) The water bag model. The ion distribution is
confined between two limiting trajectories $u_{1,2}$.}
\label{WaterBagModel}
\end{figure}
Our theory of the autoresonant evolution of the ion acoustic wave
illustrated in Figs. 1 and 2 is based on the water bag model \cite{Berk} of
the ion distribution. We assume that the distribution is constant, $%
f(u,x,t)=1/(2\Delta )$, between two trajectories $u_{1,2}(x,t)$ in phase
space and vanishes outside these trajectories (see Fig. 3). In this case,
the problem can be described by the following set of the momentum and
Poisson equations%
\begin{eqnarray}
&&u_{1t}+u_{1}u_{1x}=-\varphi _{x}  \notag \\
&&u_{2t}+u_{2}u_{2x}=-\varphi _{x}  \label{WBM} \\
&&\varphi _{xx}=\exp (\varphi +\varphi _{d})-(u_{1}-u_{2})/(2\Delta )  \notag
\end{eqnarray}%
If one defines $n(x,t)=(u_{1}-u_{2})/(2\Delta )$ and $u(x,t)=(u_{1}+u_{2})/2$%
, Eqs. (\ref{WBM}) yield
\begin{eqnarray}
&&n_{t}+(un)_{x}=0  \notag \\
&&u_{t}+uu_{x}=-\varphi _{x}-\Delta ^{2}nn_{x}  \label{FM} \\
&&\varphi _{xx}=\exp (\varphi +\varphi _{d})-n  \notag
\end{eqnarray}%
Thus, our water bag model is isomorphic to the fluid limit of the driven ion
acoustic waves with Boltzmanian electrons, the adiabatic ion pressure
scaling $p\sim n^{3}$, and $\Delta ^{2}=3\sigma ^{2}$.

Next, we observe that by defining the auxiliary potentials $\psi _{1,2}$ via
$u_{1,2}=(\psi _{1,2})_{x}$, Eqs. (\ref{WBM}) can be derived from the
variational principle with the following 3-field Lagrangian%
\begin{eqnarray}
&&L=\frac{\varphi }{2\Delta }(\psi _{1x}-\psi _{2x})-e^{\varphi +\varphi
_{d}}-\frac{1}{2}\varphi _{x}^{2}  \label{LL} \\
&&+\frac{1}{4\Delta }(\psi _{1x}\psi _{1t}-\psi _{2x}\psi _{2t})+\frac{1}{%
12\Delta }(\psi _{1x}^{3}-\psi _{2x}^{3}).  \notag
\end{eqnarray}%
This Lagrangian can be used in the Whitham's averaged variational principle
\cite{Whitham} for studying the fluid limit of the driven-chirped ion
acoustic waves. The idea is to average (\ref{LL}) over the fast oscillations
in the problem to get a new Lagrangian characterizing adiabatic modulations
of the autoresonant wave parameters and, thus, describe the slow evolution
of the system trapped in resonance with the driving wave. In studying the
aforementioned autoresonance threshold phenomenon, we limit our theory to a
weakly nonlinear evolution stage and, consequently, write the truncated
harmonic decomposition of the three potentials \cite{Whitham}: $\psi
_{i}\approx \xi _{i}+\frac{b_{i}}{k}\sin \theta +\frac{c_{i}}{2k}\sin
(2\theta )$, $\varphi =a_{0}+a_{1}\sin \theta +a_{2}\sin (2\theta )$. Here,
the amplitudes $a_{i}$, $b_{i}$, $c_{i}$ are assumed to be slow functions of
time, the wave phase $\theta $ and auxiliary phases $\xi _{i}$ (necessary
because $\psi _{i}$ enter the Lagrangian via space/time derivatives only)
are assumed to be fast, but $\theta _{x}=k$, $(\xi _{i})_{x}=\gamma _{i}\ $%
are constants (given by initial conditions), and $\theta _{t}=-\omega (t)$, $%
(\xi _{i})_{t}=-\alpha _{i}(t)$ are slow. Furthermore, we assume that the
amplitudes of the zero and second harmonics scale quadratically with the
amplitudes of the first harmonics. Then, the substitution into (\ref{LL})
(with $e^{\varphi +\varphi _{d}}$ approximated as $e^{\varphi +\varphi
_{d}}\approx 1+\varphi +\varphi ^{2}/2+\varphi ^{3}/6+\varphi
^{4}/24+\varepsilon \varphi \cos (\theta -\Phi )$, where the phase mismatch $%
\Phi =\theta -\theta _{d}$ is assumed to be slow), averaging over $\theta $
between $0$ and $2\pi $, and truncating at the fourth order in terms of the
fundamental harmonic amplitudes yields the averaged slow Lagrangian
\begin{equation}
\Lambda =\Lambda _{0}(a_{0,1,2},b_{1,2},c_{1,2};k,\omega ,\gamma
_{1,2},\alpha _{1,2})+\frac{1}{2}\varepsilon a_{1}\cos \Phi  \label{L1}
\end{equation}%
in the problem. Note that $k,\gamma _{1,2}$ in our problem are given. Taking
variations with respect to all slow amplitudes, but $a_{1}$ yields six
algebraic equations $\partial \Lambda _{0}/\partial A_{m}=0$, where $A_{m}$
represents the set $(a_{0,2},b_{1,2},c_{1,2})$. These equations allow to
eliminate all $A_{m}$ from the problem and, after the substitution back into
(\ref{L1}), obtain a new slow Lagrangian
\begin{equation}
\Lambda ^{\prime }=\Lambda _{0}^{\prime }(a_{1};\theta _{x},-\theta
_{t},(\xi _{1,2})_{x},-(\xi _{1,2})_{t})+\frac{1}{2}\varepsilon a_{1}\cos
\Phi  \label{L2}
\end{equation}%
involving the remaining amplitude $a_{1}$ and phases $\theta $ (via its
derivatives and $\Phi $) and $\xi _{1,2}$ (via their derivatives only).
Taking the variations with respect to $\xi _{1,2}$ yields two algebraic
equations $\partial \Lambda _{0}^{\prime }/\partial \alpha _{1,2}=const$,
which allow to eliminate $\alpha _{1,2}$ and obtain the final Lagrangian of
form
\begin{equation}
\Lambda ^{\prime \prime }=\Lambda _{0}^{\prime \prime }(a_{1};k,\omega )+%
\frac{1}{2}\varepsilon a_{1}\cos \Phi .  \label{L3}
\end{equation}%
The evaluation of $\Lambda _{0}^{\prime \prime }(a_{1};k,\omega )$ (to $%
O(a_{1}^{4})$) in our problem involves a lengthy algebraic manipulation,
which we performed by using Mathematica \cite{Mathematica}. Here, we present
the final result in the limit $\Delta =0$:%
\begin{equation}
\Lambda ^{\prime \prime }=\frac{1}{2}B(k,\omega )a_{1}^{2}-\frac{1}{4}%
C(k,\omega )a_{1}^{4}+\frac{1}{2}\varepsilon a_{1}\cos \Phi  \label{L4}
\end{equation}%
where $B=\frac{1+k^{2}}{2\omega ^{2}}(\omega ^{2}-\omega _{a}^{2})$, $%
C=D/\{4\omega ^{6}[k^{2}(4-16\omega ^{2})-4\omega ^{2}]\}$, and $%
D=6k^{4}\omega ^{4}-5k^{6}\omega ^{2}-2\omega ^{8}+k^{2}\omega
^{6}(1-4\omega ^{2})-4k^{8}(1+5\omega ^{2})$. Next, we use (\ref{L4}) and
take variations with respect to $\theta $ and $a_{1}$ to get
\begin{equation}
\frac{d}{dt}\left( \frac{\partial B}{\partial \omega }a_{1}^{2}\right) =-%
\frac{1}{2}\varepsilon a_{1}\sin \Phi  \label{a1}
\end{equation}%
\begin{equation}
B-Ca_{1}^{2}+\frac{\varepsilon }{2a_{1}}\cos \Phi =0  \label{phi1}
\end{equation}%
At this stage, we write $\omega =\omega _{a}+\Delta \omega $, assume
proximity to the linear resonance $\Delta \omega /\omega _{a}\ll 1$, and
expand $B$ and $C$ in Eqs. (\ref{a1}) and (\ref{phi1}) around $\omega _{a}$
to lowest significant order in $\Delta \omega $ to get%
\begin{equation}
\frac{d}{dt}\left( \frac{\partial B(\omega _{a})}{\partial \omega _{a}}%
a_{1}^{2}\right) =-\frac{1}{2}\varepsilon a_{1}\sin \Phi  \label{a11}
\end{equation}%
\begin{equation}
\Delta \omega =\left[ C(\omega _{a})a_{1}^{2}-\frac{\varepsilon }{2a_{1}}%
\cos \Phi \right] \left[ \frac{\partial B(\omega _{a})}{\partial \omega _{a}}%
\right] ^{-1}  \label{phi11}
\end{equation}%
Then, after evaluating $\partial B(\omega _{a})/\partial \omega
_{a}=k^{-1}(1+k^{2})^{3/2}$ and $C(\omega
_{a})=(4+42k^{2}+93k^{4}+81k^{6}+24k^{8})/(48k^{2})$, we have
\begin{equation}
\frac{da_{1}}{dt}=-\frac{\varepsilon k}{4(1+k^{2})^{3/2}}a_{1}\sin \Phi
\label{a12}
\end{equation}%
and (assuming passage through the linear resonance $\omega _{d}=\omega
_{a}+\alpha t$)%
\begin{equation}
\frac{d\Phi }{dt}=\Delta \omega -\alpha t=C^{\prime }a_{1}^{2}-\alpha t-%
\frac{\varepsilon k}{2(1+k^{2})^{3/2}a_{1}}\cos \Phi  \label{phi2}
\end{equation}%
where $C^{\prime }=C(\omega _{a})k(1+k^{2})^{-3/2}$. Finally, we define $%
a=\alpha ^{-1/2}C^{\prime 1/2}a_{1}$, rescaled time $\tau =\alpha ^{1/2}t$,
and rescaled driving amplitude $\mu =C^{\prime 1/2}\alpha ^{-3/4}\varepsilon
$. Note that Eqs. (\ref{a12}) and (\ref{phi2}) can be combined into a single
equation for $\Psi =a\exp (i\Phi )$:
\begin{equation}
i\Psi _{\tau }+(\left\vert \Psi \right\vert ^{2}-\tau )\Psi =\mu .
\label{NLS}
\end{equation}%
\begin{figure}[tp]
\includegraphics[width=8.5cm]{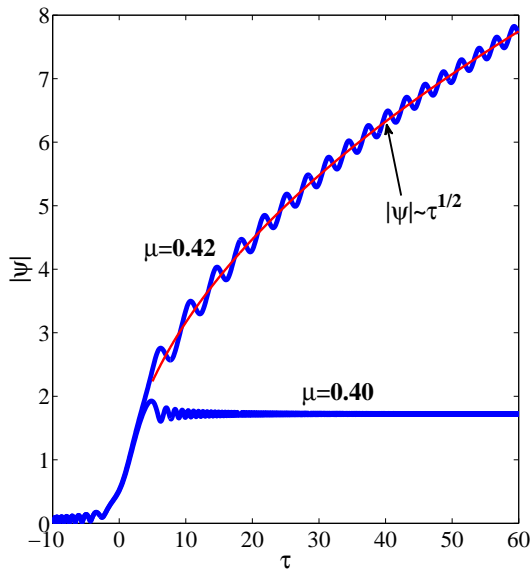}
\caption{(color online) Solutions of Eq.(\protect\ref{NLS}) for $|\protect%
\psi|$ just below ($\protect\mu=0.40$) and above ($\protect\mu=0.40$) the
threshold. The thin (red) line shows the asymptotic autoresonant solution, $|%
\protect\psi|\sim\protect\tau^{1/2}$.}
\label{abspsi}
\end{figure}
This one-parameter, nonlinear Schrodinger-type equation is characteristic to
passage through linear resonance in many dynamical systems and predicts
transition to autoresonance for $\mu >\mu _{th}=0.41$ \cite{Scholarpedia}.
We illustrate this transition phenomenon in Fig. 4, showing the evolution of
$|\psi|$ for $\mu $ just below and above $\mu _{th}$. Returning to the
original parameters in our driven ion acoustic wave problem ($\varepsilon
_{th}=0.41\alpha ^{3/4}C^{\prime -1/2}$), we obtain
\begin{equation}
\varepsilon _{th}=\frac{5.7\alpha ^{3/4}(1+k^{2})^{9/4}}{%
k^{1/2}(4+42k^{2}+93k^{4}+81k^{6}+24k^{8})^{1/2}}.  \label{thr}
\end{equation}%
The numerical solutions of Eqs. (\ref{a12}) and (\ref{phi2}) shown in Fig. 2
($\varepsilon _{th}=0.0017$ in this example) are in an excellent agreement
with Vlasov-Poisson simulations until the amplitude of the wave becomes
large in violation of our assumption of weak nonlinearity. Finally, we found
that if one uses a linear expansion $exp(\varphi +\varphi _{d})\approx
1+\varphi +\varphi _{d}$ (the usual assumption in deriving the KdV limit in
the problem) instead of the fourth order expansion used above, then $%
4+42k^{2}+93k^{4}+81k^{6}+24k^{8}$ in the denominator in the right hand side
in Eq. (\ref{thr}) should be replaced by $3(1+k^{2})^{3}(3+8k^{2})$. This
yields a significant difference in the threshold at small $k$.

In summary, we have studied autoresonant excitation of nonlinear ion
acoustic waves in the fluid approximation by passage through the linear
resonance in the problem. Weakly nonlinear Whitham's averaged variational
principle was used in the theory of the autoresonant transition, yielding a
good agreement with Vlasov-Poisson simulations. Extension of the variational
approach to include larger ion acoustic wave amplitudes, resonant kinetics,
and spatial nonuniformity effects seem to be important directions for future
research. This work was supported by the Israel Science Foundation.

\end{document}